# Stochastic Computing Implemented by Skyrmionic Logic Devices

Haoyang Zhang[&], Daoqian Zhu[&], Wang Kang*, Youguang Zhang and Weisheng Zhao*

*Fert Beijing Research Institute, BDBC, School of Microelectronics, Beihang University, Beijing, 100191, China.*

[&]These authors contributed equally to this work.

*Corresponding E-mails: wang.kang@buaa.edu.cn and weisheng.zhao@buaa.edu.cn

## Abstract

Magnetic skyrmion, topologically non-trivial spin texture, has been considered as promising information carrier in future electronic devices because of its nanoscale size, low depinning current density and high motion velocity. Despite the broad interests in skyrmion racetrack memory, researchers have been recently exploiting logic functions enabled by using the particle-like behaviors of skyrmions. These functions can be applied to unconventional computing, such as stochastic computing (SC), which treats data as probabilities and is superior to binary computing due to its simplicity of logic operation. In this work, we demonstrate SC implemented by skyrmionic logic devices. We propose a skyrmionic AND-OR logic device as a multiplier in the stochastic domain and two skyrmionic multiplexer (MUX) logic devices as stochastic adders. With the assist of voltage controlled magnetic anisotropy (VCMA), the precise control of skyrmions collision is not required in the skyrmionic AND-OR logic device, thus improving the operation robustness. In the two MUX logic devices, skyrmions can be driven by Zhang-Li torque or spin orbit torque (SOT). Particularly, we can flexibly regulate the skyrmion motion by VCMA or voltage controlled Dzyaloshinskii-Moriya Interaction (VCDMI) in the SOT case. Furthermore, 3-bit stochastic multiplier and adder are demonstrated by micromagnetic simulations. In addition, simulations in synthetic antiferromagnets (SAF) show that the performance of our skyrmionic logic gates can be optimized through advanced materials. Our work opens up perspective to implement SC using skyrmionic logic devices.

## I. INTRODUCTION

Magnetic skyrmions are topologically stable chiral structures, which are generated at the B20-type bulk material or ultra-thin ferromagnetic (FM) layer favored by Dzyaloshinskii-Moriya Interaction (DMI) [1-4]. Due to their small size, high motion velocity and low depinning current density, skyrmions are considered as promising carriers to transfer information in future electronic devices [5-9]. Over the years, much effort has been devoted in developing skyrmion racetrack memory [10-12], which requires no mechanical parts and shows great potential in advanced high-density storage. Moreover, by exploiting the particle-like behaviors of skyrmions, logic devices with high operation speed and low power consumption have also been proposed [6,7].

One of the most essential application of skyrmionic logic device is to support stochastic computing (SC) which can be realized by AND and MUX logic operations. SC is an unconventional computing method that treats data as probabilities [13-17]. A $N$-bit stochastic number (SN) bit-stream with $X$ 1s denotes a probability $P = X/2^N$, indicating the probability of observing bit 1 at a bit-stream. SC has been applied in the massively parallel computing system for its tolerance to the soft error as well as high operation speed [18]. Besides, *SN* and *X* are flexible, which brings great convenience to its applications. To date, SC has been only implemented in CMOS based stochastic circuits, which suffers from the challenges of power and area cost [19,20].

Skyrmionic logic devices emerge as a solution to these issues. The existing proposals of skyrmionic AND logic device require the precise control of skyrmions collision to execute logic operations [6,7], which relies on various effects, such as skyrmion-edge interaction [21,22] and skyrmion Hall effect [23-27]. However, recent studies have shown that the skyrmion Hall angle varies with skyrmion velocity in the presence of defects and thermal fluctuations, therefore increasing the difficulty to accurately control the skyrmion trajectory [28,29]. In addition, recent studies show that skyrmions below 10 nm cannot be stabilized in ferromagnetic multilayers without external magnetic fields because of the demagnetization field while small skyrmions in ferromagnetic thin films suffer from thermal stability issue [30,31]. As a consequence, it is necessary to demonstrate skyrmionic gates in ferrimagnets [32] or synthetic antiferromagnets (SAF) [33-35], which are promising to host sub-10 nm skyrmions at room temperature. Since skyrmion Hall effect is (largely) suppressed in such systems [34], new schemes to realize skyrmion logic gates are in demand. Furthermore, no skyrmionic MUX logic device has been proposed yet.

In this letter, we propose a skyrmionic AND-OR logic device and two types of skyrmionic MUX logic devices regarding the skyrmion motion manner to implement SC. In the skyrmionic AND-OR logic device which acts as a stochastic multiplier [36], skyrmions are driven by spin orbit torque (SOT) [37,38] and guided by voltage controlled magnetic anisotropy (VCMA) effect [39,40]. As a consequence, it is not necessary to accurately control skyrmions collision to implement logic operations and this device can tolerate the thermal diffusion under $T = 250$ K. In the skyrmionic MUX logic devices operating as a stochastic adder [36], skyrmions can be moved by Zhang-Li torque [41] or SOT. VCMA or voltage controlled DMI (VCDMI) [42] is used to dynamically modify the energy landscape to regulate skyrmion motion in such devices. Based on the proposed logic devices, 3-bit SN stochastic multiplier and stochastic adder are further confirmed by micromagnetic simulations. We also demonstrate that our proposed skyrmionic logic devices can work in the SAF, where no skyrmion Hall effect exists and small magnetic skyrmions can be stabilized at room temperature. The performances of the proposed skyrmionic logic devices in FM and SAF are analyzed, which shows broad prospects of the proposed skyrmionic logic devices on implementing SC.

## II. Skyrmionic AND-OR Logic Device

In the proposed structure shown in Fig. 1(a), the skyrmionic AND-OR device is mainly composed of three parts, including heavy metal (HM) layer, ferromagnetic (FM) layer and two voltage electrode gates. The HM layer, which is beneath the FM layer (not shown in the figure), is used for the flow of driving current to drive skyrmion from left side to right side by spin Hall effect [43]. We apply positive voltages on electrode gate 1 ($V_{g1}$) and 2 ($V_{g2}$) (see Fig.1(b)), increasing the potential energy of the FM layer beneath $V_{g1}$ and $V_{g2}$ to guide the skyrmion motion. The effect of voltage on perpendicular magnetic anisotropy (PMA) can be calculated by the equation $K_g = K_u + \xi \cdot V_g / (d \cdot h)$, where the VCMA coefficient $\xi$ is set as 48 fJ V$^{-1}$ m$^{-1}$ according to previous reports [40]. $d$ denotes the thickness of the insulator layer between the electrode gates and FM layer, and $h$ is the thickness of FM layer, which are both set as 1 nm in our simulations. Precisely, if a +2 V voltage is applied on one electrode gate, the PMA of FM layer beneath it will increase from 0.8 MJ m$^{-3}$ to 0.896 MJ m$^{-3}$, creating an energy barrier blocking the skyrmions from crossing this region (see Appendix A for other simulation details).

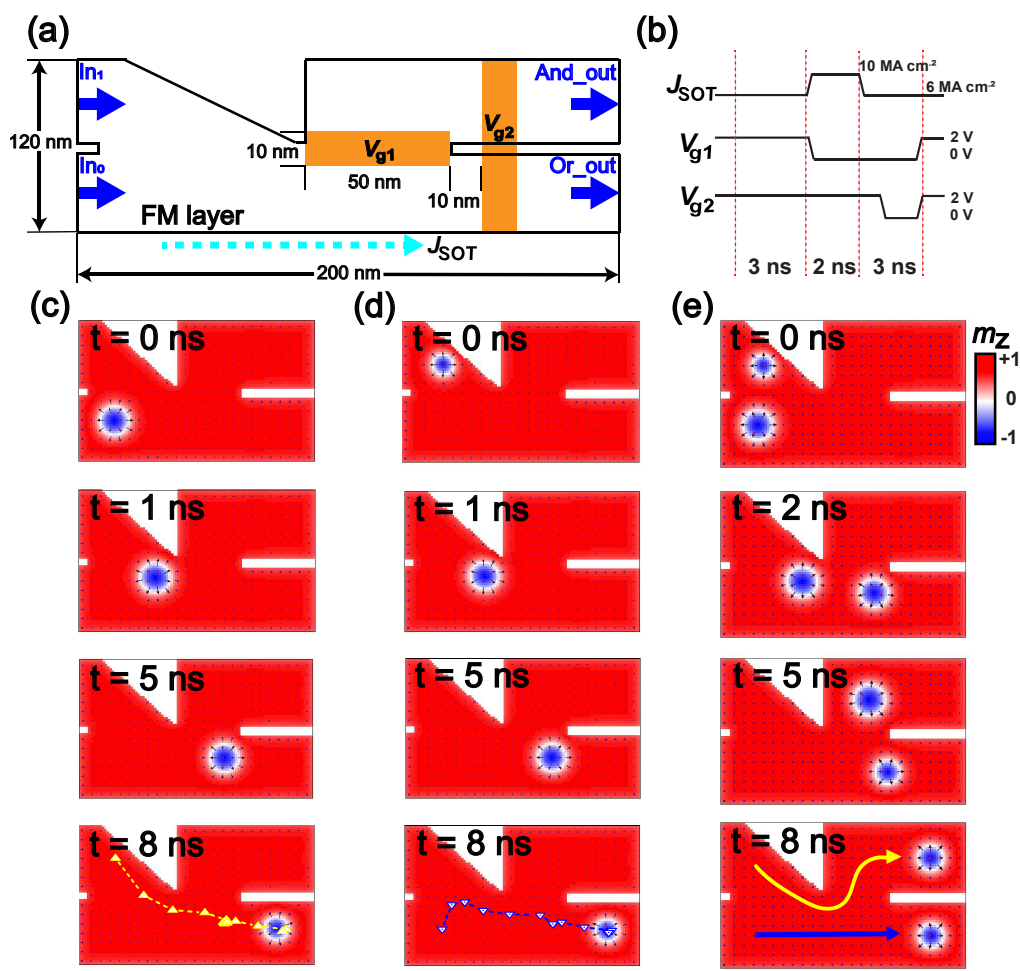

FIG. 1 Skyrmionic AND-OR logic device at $T = 0$ K. (a) Device structure with two input lanes and two output lanes. The two electrode gates ($V_{g1}$, $V_{g2}$) are designed to modulate skyrmion motion through VCMA effect. (b) Profile of driving current density ($J_{SOT}$) and applied positive voltages to achieve device function. (c-e) Three cases of inputs and their corresponding simulation processes and outputs. In the snapshots of micromagnetic simulation, color red represents spin up ($m_z = +1$), blue represents spin down ($m_z = -1$) and white represents spin that horizonal to the plane ($m_z = 0$). Skyrmion trajectories are indicated in the magnetization snapshots at $t = 8$ ns.

Fig. 1(b) shows the profile of SOT current density ($J_{SOT}$) and electrode voltages to achieve device function. A voltage of +2V is dynamically applied on $V_{g1}$ and $V_{g2}$ to guide skyrmion motion. Fig.1(c-e) demonstrates the top-viewed simulation snapshots under three different inputs. The data "1" or "0" is encoded by the presence or absence (FM background) of a skyrmion. In the circumstance with only one skyrmion input, this skyrmion will enter OR lane and be blocked at a position near $V_{g2}$ under the joint effect of VCMA, $J_{SOT}$ = 6 MA cm$^{-2}$ for 3 ns and repulsion from the edge [21,22]. After the positive voltage on $V_{g2}$ is removed, the skyrmion finally outputs from the OR lane. Therefore, the OR and AND lane generate an output of "1" and "0", respectively. In the case with two skyrmion inputs, the first skyrmion (from In$_0$) stops near V$_{g2}$, similar to the aforementioned behavior. The subsequent skyrmion (from In$_1$) is then blocked because of the repulsion from the first skyrmion and the energy potential caused by $V_{g1}$. At $t$ = 3 ns, the voltage on $V_{g1}$ is set to 0 and $J_{SOT}$ increases to 10 MA cm$^{-2}$. Thus the subsequent skyrmion will be driven to the AND lane by a combined effect of skyrmion Hall Effect [23-27], skyrmion-skyrmion repulsion and skyrmion-edge repulsion while the first skyrmion moves out from OR lane after the voltage on $V_{g2}$ returns to 0. In this case, both OR and AND lane give rise to an output of data "1". Obviously, AND-OR logic function is realized in our proposed device. Note that during these operations, it is not necessary to precisely control the skyrmion-skyrmion collision. However, in the precious proposals [6,7], this is required to realize logic function, which will suffer from instability in the presence of thermal effect and pinning. We further verify the function of the proposed AND-OR Logic device at $T$ = 250 K (see Appendix B for details), indicating the operation robustness of our device.

Moreover, we designed a 3-bit (length of stochastic bit stream of 8) stochastic multiplier, using the proposed skyrmionic logic AND-OR device as the stochastic multiplier cell. Fig. 2(a) shows the schematic of multiplication computation of 4/8 × 6/8 implemented by an AND logic gate, which can be physically achieved in our proposed skyrmionic device, as exhibited in Fig. 2(b). The AND function is implemented using the above skyrmionic logic AND-OR device while the inputs and outputs can be stored in the corresponding elongated lanes. Notches at the lanes are used to synchronize the input and output skyrmions, and divide each lane into 8 regions (yellow regions in Fig. 2(b)). Through periodically increasing the current to $J_{SOT} = 20$ MA cm$^{-2}$ for 0.6 ns, skyrmions can move to the region of the next data bit [7]. Otherwise, skyrmions will be blocked in their original area. The operation timing diagram of the AND-OR logic operation in one cycle is shown in Fig. 2(c). Compared to the timing diagram in Fig. 1(b), an additional 0.6 ns wide high current pulse is provided every 19.6 ns to synchronize the skyrmions, which means that the whole operation time for 3-bit multiplication computing is about 156.8 ns. In our simulations, input skyrmions are initialed at the left racetracks. There are four and six skyrmions distributed in the two input lanes, denoting two probabilities of $P_1 = 4/8$ and $P_0 = 6/8$, respectively. Note that skyrmion nucleation time was not considered in our simulations, as recent studies have shown that skyrmions can be generated in only ~0.6 ns by SOT pulses [9]. More interestingly, the input skyrmion stream can also be provided through a reshuffler device with a skyrmionic reservoir [44], in which the skyrmions are generated asynchronously. Therefore, the skyrmion nucleation time will not bound the clock speed of whole system. After eight cycles, three skyrmions output from the AND lane, as shown in Fig. 2(d), which illustrates an output probability $P_{out} = 3/8$ and the capableness of our device to perform stochastic multiplication.

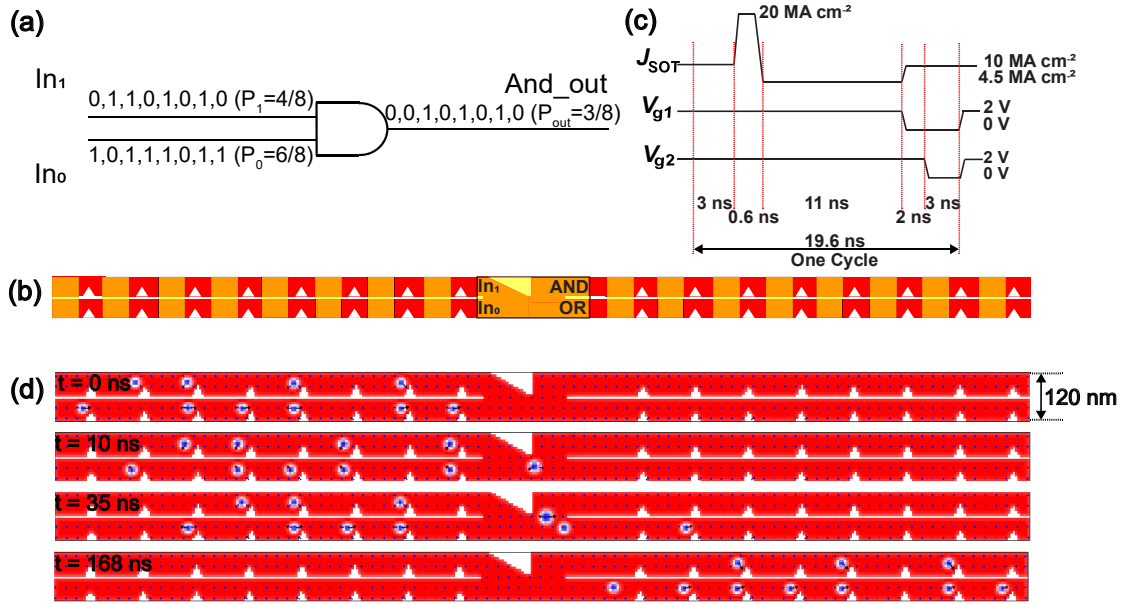

FIG. 2 3-bit Skyrmionic AND-OR logic device used as a stochastic multiplier. (a) Schematic of exact multiplication computation of $4/8 \times 6/8 = 3/8$. A logic AND gate is required during the operation. (b) Schematic of the proposed 3-bit skyrmionic stochastic multiplier. Notches are used to divide each lane into 8 regions and synchronize the motion of skyrmions. (c) Profile of driving current density ($J_{SOT}$) and electrode voltages in an operation cycle. (d) Snapshots of magnetization configuration of the 3-bit stochastic multiplier at selected times. At $t = 0$ ns, the skyrmions distributed in the two input lanes represent probabilities $P_1 = 4/8$ and $P_0 = 6/8$, respectively. After 168 ns, AND lane will output the binary sequence of "0, 0, 1, 0, 1, 0, 1, 0", i.e., probability $P_{out} = 3/8$, which denotes the multiplication result of two inputs.

### III. Skyrmionic MUX Logic Device

SC addition operation can be performed by a multiplexer (MUX) [36] where two input bit streams ($In_1$, $In_0$) are selected by a select signal S with a probability of 0.5, thus the output bit stream owns a probability of half the sum of two inputs. Considering the feature of MUX and the particle-like behaviors of skyrmions, we propose two types of skyrmionic MUX logic devices regarding the skyrmion motion manner.

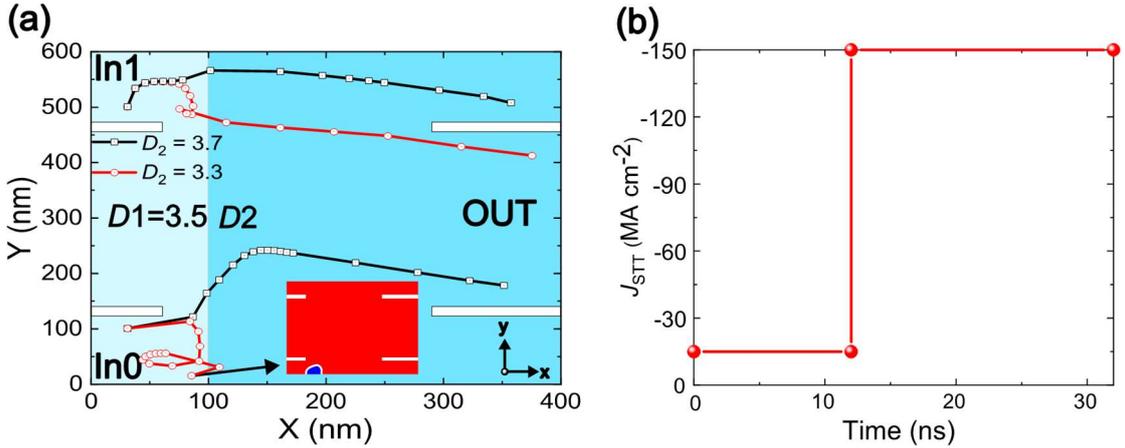

FIG. 3 Skyrmionic MUX logic device driven by Zhang-Li torque. (a) Schematic of the proposed device, where the DMI in the left panel ($D_1$) and right panel ($D_2$) is different. There are two input lanes ($In_0$, $In_1$), one output lane (OUT) to export the selected data. Trajectories of skyrmions are indicated by the black and red symbols when $D_2 = 3.35$ or 3.65 mJ m$^{-2}$ while $D_1$ is fixed as 3.5 mJ m$^{-2}$. (b) Profile of driving in-plane polarized current $J_{STT}$. Note that damping $\alpha = 0.1$ is used in the simulations of (a) to obtain large enough skyrmion deflection.

Fig. 3(a) shows one scheme to implement the MUX logic, which is based on the deflection of skyrmions at an interface between regions with different DMI. The DMI constant $D_1$ of the left side is set as 3.5 mJ m$^{-2}$ while $D_2$ of the right plane varies from 3.35 to 3.65 mJ m$^{-2}$. The skyrmion deflection direction depends on the relative magnitude of $D_1$ and $D_2$. Specifically, in our settings, if $D_2 > D_1$, skyrmions will be deflected along +$y$ direction. Therefore, skyrmions from In$_0$ will enter the OUT lane while skyrmions from In$_1$ will output from C$_1$ lane, which means data from In$_0$ is selected. In contrast, if $D_2 < D_1$, skyrmions from In$_1$ will enter OUT lane, while skyrmion from In$_0$ will crush at the lower edge. By regulating $D_2$ using the VCDMI effect according to an electronical select signal S to select two skyrmionic input signals, MUX logic can be implemented in our device. Note that when skyrmions cross the interface where the DMI changes, the applied in-plane polarized current (CIP) $J_{STT}$ maintains 15 MA cm$^{-2}$ (see Fig. 3(b)). If the current density is too high, skyrmion will quickly go through the interface, leading to a small displacement along $y$ direction, not enough to realize the function.

However, this scheme has several shortcomings. First, the stability and power efficiency will be affected by the crush of skyrmions. Second, deflection toward –$y$ direction at the interface can only be achieved when skyrmions are driven by Zhang-Li torque (see Appendix C for the simulation results of SOT driven case and Discussion for explanation), which consumes more energy and time than the SOT driven case. In addition, the skyrmion displacement in the $y$ direction is inversely proportional to the damping [45], which indicates that damping should be relatively low (damping $\alpha = 0.1$ in our simulations) to enable an enough displacement. But in the widely studied Pt/Co system, damping is usually larger than 0.1 [46].

Therefore, we propose another scheme of skyrmionic MUX logic device where skyrmions are driven by SOT. Besides, there are two copy lanes (C$_1$, C$_0$) outputting the data that is not selected by S. As shown in Fig. 4(a), an oblique interface is introduced to guide skyrmion motion. Three electrode gates ($V_{G1}$, $V_{G2}$, $V_{G3}$) are used to lower the DMI value of the FM layer beneath the electrode gates through VCDMI effect [40]. $V_{G1}$ and $V_{G3}$ are always applied with the same voltage, used to select the input skyrmion from In$_1$. By contrast, a voltage different from $V_{G1}$ and $V_{G3}$ is applied on $V_{G2}$ to select skyrmion from In$_0$. The select signal S can be produced electronically by a stochastic number generator (SNG) [47], where the probability of "1" is 0.5, as done in our simulations. S can also be produced by skyrmionic devices or superparamagnetic magnetic tunnel junction based SNG [48,49]. Recently, energy-efficient SC has been demonstrated via superparamagnetic magnetic tunnel junctions [50]. The optimal approach is to utilize a stochastic skyrmion stream which already exists in the stochastic computing system, like the output bitstreams in Fig. 2(d) to achieve "all-skyrmionic" implementation, which, however, needs further investigation. When the random number from S is 0, $D_2$ will decrease to 3.2 mJ m$^{-2}$, creating a high energy barrier to guide skyrmion from In$_0$ to enter the OUT lane. In the meantime, $D_1$ and $D_3$ remain unchanged (default value $D$ = 3.5 mJ m$^{-2}$). Therefore, skyrmions from In$_1$ will output from C$_1$. When the random number is 1, only $V_{G1}$ and $V_{G3}$ are selected, lowering $D_1$ and $D_3$ to 3.2 mJ m$^{-2}$ simultaneously. Therefore, skyrmions from In$_1$ eventually move to the OUT lane while skyrmions from In$_0$ are blocked by $V_{G3}$ and enter C$_0$ lane. The trajectories of skyrmions under different circumstance are explicitly shown in Fig. 4(a).

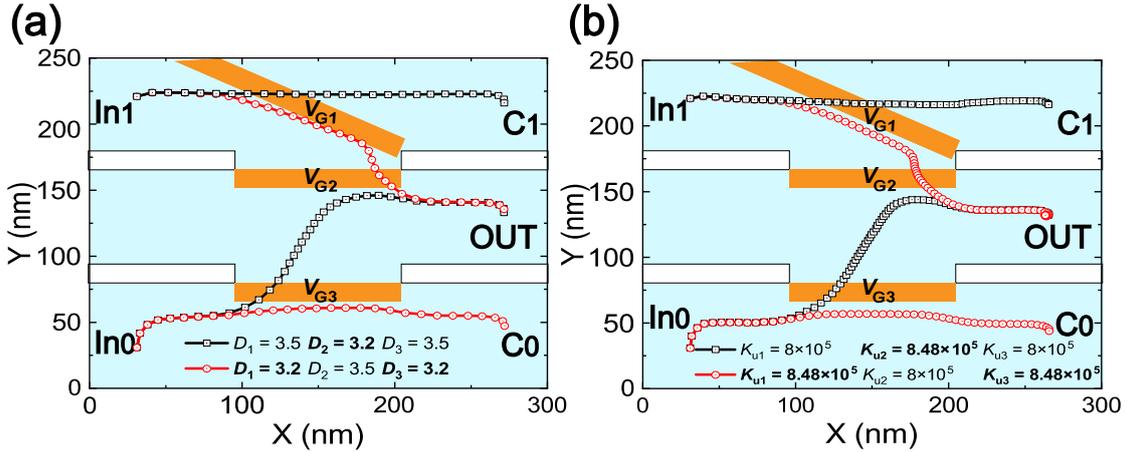

FIG. 4 Skyrmionic MUX logic devices driven by SOT. Schematic of the proposed device where energy landscape is dynamically changed through (a) VCDMI effect (b) VCMA effect. Three electrode gates ($V_{G1}$, $V_{G2}$, $V_{G3}$) and an oblique interface are introduced to guide skyrmion motion. Trajectories of the skyrmions are indicated by the blue and red symbols when selecting data from different lanes. $J_{SOT}$ applied in (a) and (b) is 6.1 MA cm$^{-2}$ and 4.0 MA cm$^{-2}$, respectively.

During the operation, the 0.3 mJ m$^{-2}$ difference of DMI can sustain a max current density of $J_{SOT}$ = 6.1 MA cm$^{-2}$. If the current is higher, skyrmions from In$_1$ will go through the $V_{G1}$ region and enter the C$_1$ lane, leading to the wrong logic output. Except for VCDMI, VCMA effect can also be used to change the local potential energy. Fig. 4(b) shows the schematic of a skyrmionic MUX device using VCMA. Similar to Fig. 4(a), three voltage gates are placed to modify the energy landscape in the device. In our simulations, we apply a +1 V voltage on the electrode gates to increase the local PMA energy density from 0.8 MJ m$^{-3}$ to 0.848 MJ m$^{-3}$. In this case, the max current density is $J_{SOT}$ = 3.0 MA cm$^{-2}$, otherwise skyrmions will enter into wrong lanes. Materials with higher VCMA coefficient of higher voltages can be used to improve the device speed.

Based on the MUX device using VCMA, we further design a 3-bit stochastic adder. Fig. 5(a) is the schematic of exact addition computation of 1/2 · (7/8 + 3/8) performed by a MUX device. The structure of the proposed 3-bit skyrmionic stochastic adder is shown in Fig. 5(b). Following the timing diagram in Fig. 5(c), the proposed skyrmionic MUX logic device can select skyrmions from In$_1$ or In$_0$ depending on the sequence S produced by SNG. Similar to the 3-bit stochastic multiplier, a high current $J_{SOT}$ = 10 MA cm$^{-2}$ for 2.2 ns is used every 17.2 ns to enable skyrmions cross the notches and synchronize the skyrmions. Therefore, the time to complete the 3-bit MUX logic is 137.6 ns. Fig. 5(d) shows the snapshots of magnetization configuration at selected times. We can see that there are five skyrmions on the OUT lane, denoting a probability $P_{out}$ = 5/8, which is exactly the half sum of the two input probabilities $P_1$ = 7/8 and $P_0$ = 3/8.

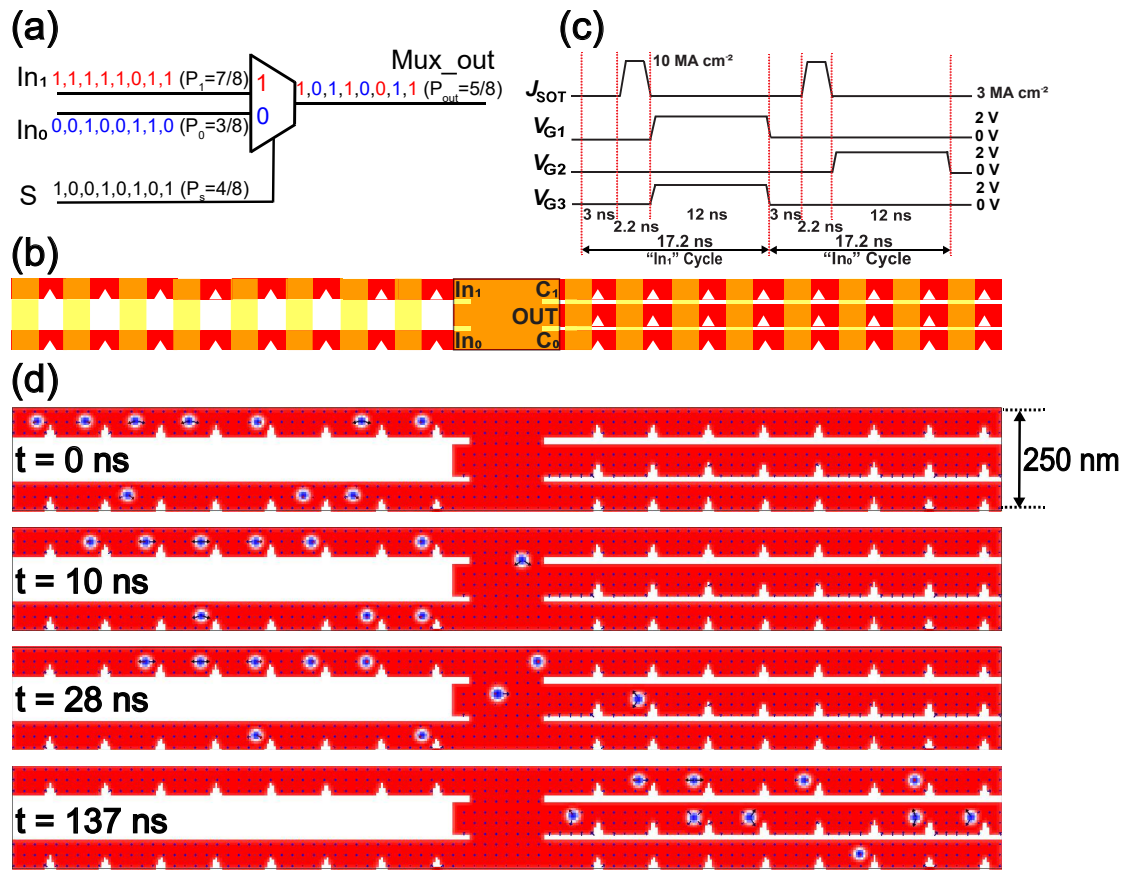

FIG. 5 3-bit Skyrmionic MUX logic device used as a stochastic adder. (a) Schematic of exact addition computation of 1/2 · (7/8 + 3/8) = 5/8 implemented by a MUX device. (b) Schematic of the proposed 3-bit stochastic adder. (c) Profile of driving current density ($J_{SOT}$) and applied voltages on three electrode gates ($V_{G1}$, $V_{G2}$ and $V_{G3}$) in an operation cycle (d) Snapshots of magnetization configuration of the 3-bit stochastic multiplier at selected times. At $t = 0$ ns, the skyrmions distributed in the two input lanes represent probabilities $P_1 = 7/8$ and $P_0 = 3/8$, respectively. After 137 ns, the OUT lane will output the binary sequence of "1, 0, 1, 1, 0, 0, 1, 1", i.e., probability $P_{out}$ = 5/8, which denotes the half sum result of two inputs.

## IV. Discussion

### A. Skyrmion dynamics at an interface where DMI changes

To understand the skyrmion dynamics in the two proposed MUX logic devices, we investigate the trajectory of skyrmions driven by Zhang-Li torque and SOT based on Thiele equation which assumes that skyrmions are rigid during the steady motion.

We first analyze the skyrmion dynamics driven by Zhang-Li torque. The Thiele equation for this case reads [51]:

$$\boldsymbol{G} \times (\boldsymbol{v} - \boldsymbol{u}) - \alpha D \boldsymbol{v} + \boldsymbol{F_i} = 0 \quad (1)$$

Here, $\boldsymbol{G} = (0, 0, G)$ is the gyromagnetic coupling vector, where $G = \frac{M_s t_F}{\gamma} 4\pi Q$ with saturation magnetization $M_s$, FM layer thickness $t_F$, gyromagnetic ratio $\gamma$, and skyrmion number $Q = -\frac{1}{4\pi} \iint \boldsymbol{m} \cdot (\frac{\partial \boldsymbol{m}}{\partial x} \times \frac{\partial \boldsymbol{m}}{\partial y}) dS$; $\boldsymbol{v} = (v_x, v_y, 0)$ is skyrmion motion velocity; $\boldsymbol{u}$ is a vector along the electron motion direction with amplitude $u$; $\alpha$ is the damping constant; $D = \frac{M_s t_F}{\gamma} \iint (\frac{\partial \boldsymbol{m}}{\partial x})^2 dS$ is the dissipative force; $\boldsymbol{F_i} = -\nabla V(\boldsymbol{r})$ is the force induced by the interface where DMI changes with $V(\boldsymbol{r})$ denoting the system energy when a skyrmion locates at position $\boldsymbol{r}$. Assume that we apply an in-plane current along $-x$ direction and the interface is exactly along $y$ direction as shown in Fig. 3(a), the solution to Eq. (1) is given by:

$$\begin{cases} v_x = \dfrac{\alpha D F_i + G^2 u}{G^2 + (\alpha D)^2} \\ v_y = \dfrac{G(F_i - \alpha D u)}{G^2 + (\alpha D)^2} \end{cases} \quad (2)$$

where $F_i$ denotes the magnitude of the force $\boldsymbol{F_i}$.

If $D_1 = D_2$ or skyrmions are far away enough from the interface, we can set $F_i = 0$ in Eq. (2). As a consequence, we obtain $v_x = \frac{G^2 u}{G^2+(\alpha D)^2}$ and $v_y = \frac{-\alpha D G u}{G^2+(\alpha D)^2}$. This result illustrates that skyrmions will gain a velocity along $+x$ direction in this case, which is the feature of Zhang-Li torque driven domain wall dynamics. In contrast, whether skyrmions move toward $+y$ or $-y$ direction depends on the sign of skyrmion number $Q$. In our simulations, we get $Q = 1$ because the magnetization of the background FM layer points along $+z$ direction. Therefore, $v_y$ is negative, which agrees with the trajectories shown in Fig. 3(a). Further, the skyrmion Hall angle $\theta_{sk}$, which describes the deviation of skyrmions from $x$ direction, is given by $\theta_{sk} = \text{atan}\left(\frac{v_y}{v_x}\right) = -\text{atan}\left(\frac{\alpha D}{G}\right)$. If the skyrmion radius $R$ is larger than the domain wall width $\Delta$, $\theta_{sk}$ can be well simplified to $-\text{atan}\left(\frac{\alpha R}{2\Delta}\right)$ [52]. Taking the parameters used in Fig. 3(a), $\theta_{sk}$ is estimated as $-16°$, corresponding well to the simulation results. When $D_1 \neq D_2$ and skyrmion are near the interface, $F_i$ cannot be left out in Eq. (2). As illustrated in Fig. 7(c), $F_i$ is positive if $D_1 < D_2$ and vice versa. From the expression of $v_y$, we can see that the skyrmion deflection can be changed by varying the magnitude of $F_i$ and $\alpha D u$. Once $D_1 < D_2$ and $F_i > \alpha D u$, skyrmions will move toward $+y$ direction. Otherwise, skyrmions still bend over $-y$ direction. The first proposed MUX logic device is exactly based on this property.

In contrast, the Thiele equation for the SOT driven case reads [43]:

$$\boldsymbol{G} \times \boldsymbol{v} - \alpha D \boldsymbol{v} + \boldsymbol{F}_{SHE} + \boldsymbol{F}_i = 0 \quad (3)$$

where $\boldsymbol{F}_{SHE} = F_{SHE}\hat{x}$ is the force induced by SOT with magnitude $F_{SHE}$, the sign of which depends on the spin polarization direction $\sigma$. Solving Eq. (3), we find,

$$\begin{cases} v_x = \dfrac{\alpha D (F_{SHE} + F_i)}{G^2 + (\alpha D)^2} \\ v_y = \dfrac{G(F_{SHE} + F_i)}{G^2 + (\alpha D)^2} \end{cases} \quad (4)$$

If $D_1 < D_2$, $F_{SHE}$ and $F_i$ are both positive, thus the skyrmion Hall angle $\theta_{sk} = \text{atan}\left(\frac{v_y}{v_x}\right) = \text{atan}\left(\frac{G}{\alpha D}\right)$, which means that skyrmions can only bend toward $+y$ direction when approaching the interface from $In_0$ or $In_1$. According to the expression of $\theta_{sk}$, the skyrmion Hall effect should be remarkable in the SOT driven case, agreeing well with the trajectories in Fig. 4(a). If $D_1 \geqslant D_2$, $F_{SHE} + F_i = 0$ can be achieved when skyrmions gradually approach the interface. Thereafter skyrmions remain static, as indicated by Fig. 7(a). Therefore, in the MUX logic device displayed in Fig. 3(a) where the interface is along y direction, only Zhang-Li torque can be used to realize the function.

If an oblique interface is introduced, as is the case in Fig. 4(a), $\boldsymbol{F_i}$ can be expressed as $\boldsymbol{F_i} = (F_{ix}, F_{iy}, 0)$. Therefore, the solution to Eq. (3) becomes:

$$\begin{cases} v_x = \dfrac{\alpha D(F_{SHE} + F_{ix}) - GF_{iy}}{G^2 + (\alpha D)^2} \\ v_y = \dfrac{G(F_{SHE} + F_{ix}) + \alpha D F_{iy}}{G^2 + (\alpha D)^2} \end{cases} \quad (5)$$

To guide skyrmions move along the interface, we have $F_{SHE} > 0$, $F_{ix} < 0$ and $F_{iy} < 0$, i.e., $D_1 > D_2$. Assume that the interface encloses angle $\delta$ with the $-y$ direction, we have $k = \tan(\delta) = -\dfrac{v_x}{v_y} = \dfrac{F_{iy}}{F_{ix}}$. Subsequently, we can obtain:

$$F_i = -\dfrac{\alpha D + kG}{\alpha D \sqrt{1+k^2}} F_{SHE} \quad (6)$$

Therefore, we can guide skyrmion motion along a designed interface when a proper SOT current is applied. Since $F_i$ can also be generated at the interface where PMA changes, VCMA effect can be used to realize similar function, as shown in Fig. 4(b). Note that if the applied SOT current is too strong, $F_i$ provided by the interface may fail to satisfy Eq. (6). Under this circumstance, skyrmions can go through the electrode gated region, resulting in a wrong output.

### B. Performance evaluation of SC implemented by skyrmionic logic devices

As shown above, we can implement SC by the skyrmionic logic devices. We then evaluate the performance of the skyrmionic logic devices. The energy consumption for the elementary skyrmionic AND-OR and MUX logic gates are 7.19 fJ and 7.29 fJ, respectively (see Appendix D for the energy calculation method), while the time delay is 8 ns and 15 ns, respectively. The synchronization clock of the 3-bit stochastic multiplier and adder is 19.6 ns and 17.2 ns, respectively, which indicates a whole operation time about 20 μs given that the stochastic number is 10. From the point of computing, skyrmion life time of approximate 1 ms may be sufficient. But it is still necessary to enhance skyrmion thermal stability if the stochastic number increases or we utilize these skyrmions for storage in the meantime. The device performance can be further optimized via adopting advanced materials like ferrimagnets [32] and SAF [33], where skyrmion velocity can be much enhanced. For example, our simulations in SAF (see Appendix E) show that the energy consumption for the proposed skyrmionic logic devices can be reduced by about 2 times at nearly the same operation speed. It is also confirmed that the proposed skyrmionics logic gates can be achieved without skyrmion Hall effect, different from previous proposals. Regarding the area, the device area depends on the skyrmion size. Using the parameters in our simulations, which are used to mimic the Pt/Co system, the diameter of skyrmion is ~20 nm. Though skyrmions of this size may encounter thermal stability issue in ferromagnetic thin film, for example, researchers only identified skyrmions of ~100 nm in Pt/Co/MgO at room temperature [53], skyrmions as small as 20-30 nm in radius have been found in Pt/Co based synthetic antiferromagnets even at $T = 420$ K [33], where the thermal stability of skyrmions is enhanced by increasing their volume while the demagnetization field is cancelled in such a system [31,33-35]. Micromagnetic simulations further predict the potential of such a system to host sub-10 nm skyrmions at room temperature [33]. The parameters adopted in this manuscript can correspond to a practical material system to realize skyrmionic logic gates. In the meantime, recent study has demonstrated skyrmions with a diameter as small as ~1 nm [3]. Though such small skyrmions have only been identified at low temperature, the huge progress made during the past ten years to stabilize small skyrmions at room temperature, even below 10 nm [32], indicates that smaller skyrmions can be anticipated and there is still much room/potential for the improvement in device area and power consumption. In addition, owing to the nonvolatile property of skyrmion, there is no static power consumption for skyrmionic logic devices while that for CMOS AND gate is ~0.9 nW [54], resulting in unneglectable energy waste in the static status. Apart from driving skyrmions in an electrical manner, magnonic [55] and/or caloritronic [56] driving methods may open up feasible avenues for our work to applications, which is beyond the scope of this paper and the details will be omitted here. Therefore, although more effort is required, our proposed skyrmionic logic design demonstrates broad prospects as an alternative to implement SC.

## V. Conclusion

In this work, we propose a skyrmionic AND-OR logic device and two types of skyrmionic MUX logic devices to implement SC. Thanks to the energy landscape designed by VCMA or VCDMI, precise control of skyrmion-skyrmion collision is not required in our devices, thus enabling high operation robustness. Based on these devices, we demonstrate the 3-bit stochastic multiplier and adder. Performance evaluation shows that our proposed skyrmionic logic devices are promising to execute SC in terms of energy consumption, time delay and area with advance in materials. Our work unfolds the great potential of skyrmionic logic devices to implement SC.

## Acknowledgements

The authors gratefully acknowledge the Beijing Natural Science Foundation (Grants No. 4202043), National Natural Science Foundation of China (Grants No. 61871008, No. 61627813 and No. 1602013), the International Collaboration Project B16001, and the National Key Technology Program of China 2017ZX01032101.

## Appendix A: Simulation Methods

Micromagnetic simulations were performed by utilizing the GPU-accelerated simulation software Mumax3 [57]. The dynamics at each site are depicted by the following Landau-Lifshitz-Gilbert (LLG) equation [4,43]:

$$\frac{\partial \boldsymbol{m}}{\partial t} = -\gamma\mu_0(\boldsymbol{m} \times \boldsymbol{H}_\text{eff}) + \alpha\left(\boldsymbol{m} \times \frac{\partial \boldsymbol{m}}{\partial t}\right) + \tau_\text{SOT} + \tau_\text{CIP}$$

where $H_\text{eff}$ is the effective field, including the contributions from the anisotropy field, exchange field, DMI field, demagnetization field and thermal field (when $T \neq 0$ K).

$m = \frac{M}{|M_s|}$ is the reduced magnetization with $M_s$ the saturation magnetization, $\alpha$ is the damping constant, $\gamma$ is the gyromagnetic ratio and $\mu_0$ is the vacuum magnetic permeability. $\tau_{SOT} = \gamma \frac{J_{SOT}\theta_{SH}\hbar}{2edM_s}(m \times \sigma \times m)$ denotes the SOT exerted on the magnetization when applying a SOT current $J_{SOT}$ with spin polarization direction $\sigma$, spin Hall angle $\theta_{SH}$, elementary charge $e$ and FM layer thickness $d$ and reduced Plank constant $\hbar$ [43]. $\tau_{CIP} = u(m \times \frac{\partial m}{\partial x} \times m)$ is the exerted Zhang-Li torque when an in-plane polarized current is applied, where $u = \gamma \frac{J_{STT}P\hbar}{2eM_s}$ with current density $J_{STT}$ and spin polarization $P$ [41,43].

Unless specified, the following material parameters are adopted in our simulations: exchange stiffness $A$ = 15 pJ m$^{-1}$, Gilbert damping $\alpha$ = 0.3, saturation magnetization $M_s$ = 580 kA m$^{-1}$, spin polarization $P$ = 0.4, spin Hall angle $\theta_{SH}$ = 0.4, PMA constant of the FM layer $K_u$ = 0.8 MJ m$^{-3}$, DMI constant $D$ = 3.5 mJ m$^{-2}$ and VCMA coefficient $\xi$ = 48 fJ V$^{-1}$ m$^{-1}$ [34,35,40]. The simulation area is divided into a cubic grid of 2 nm × 2 nm × 1 nm.

## Appendix B: Skyrmionic AND-OR functions performed at $T$ = 250 K

We enlarge the area of device and adjust timing diagram (see Fig. 6(a-b)) to enable the fluctuation of the skyrmions because of the high operation robustness of skyrmions in a wider racetrack [58]. Fig. 6(c-e) shows the simulation results with different inputs. The volume of skyrmion dramatically increases and fluctuating a lot at $T$ = 250 K, which limits the current density applied to drive the skyrmions. Basically, $J_{SOT}$ = 4.2 MA cm$^{-2}$ is applied to drive skyrmion motion but a lower current is used ($J_{SOT}$ = 2.9 MA cm$^{-2}$) when the second skyrmion moves to the AND lane. This is because the second skyrmion is more likely to crush at the edge while it fluctuates at $T$ = 250 K. The operation time becomes longer because of the larger size of the device as well as the lower driving current. But still, the AND-OR function can be achieved, indicating the high operation robustness of our device.

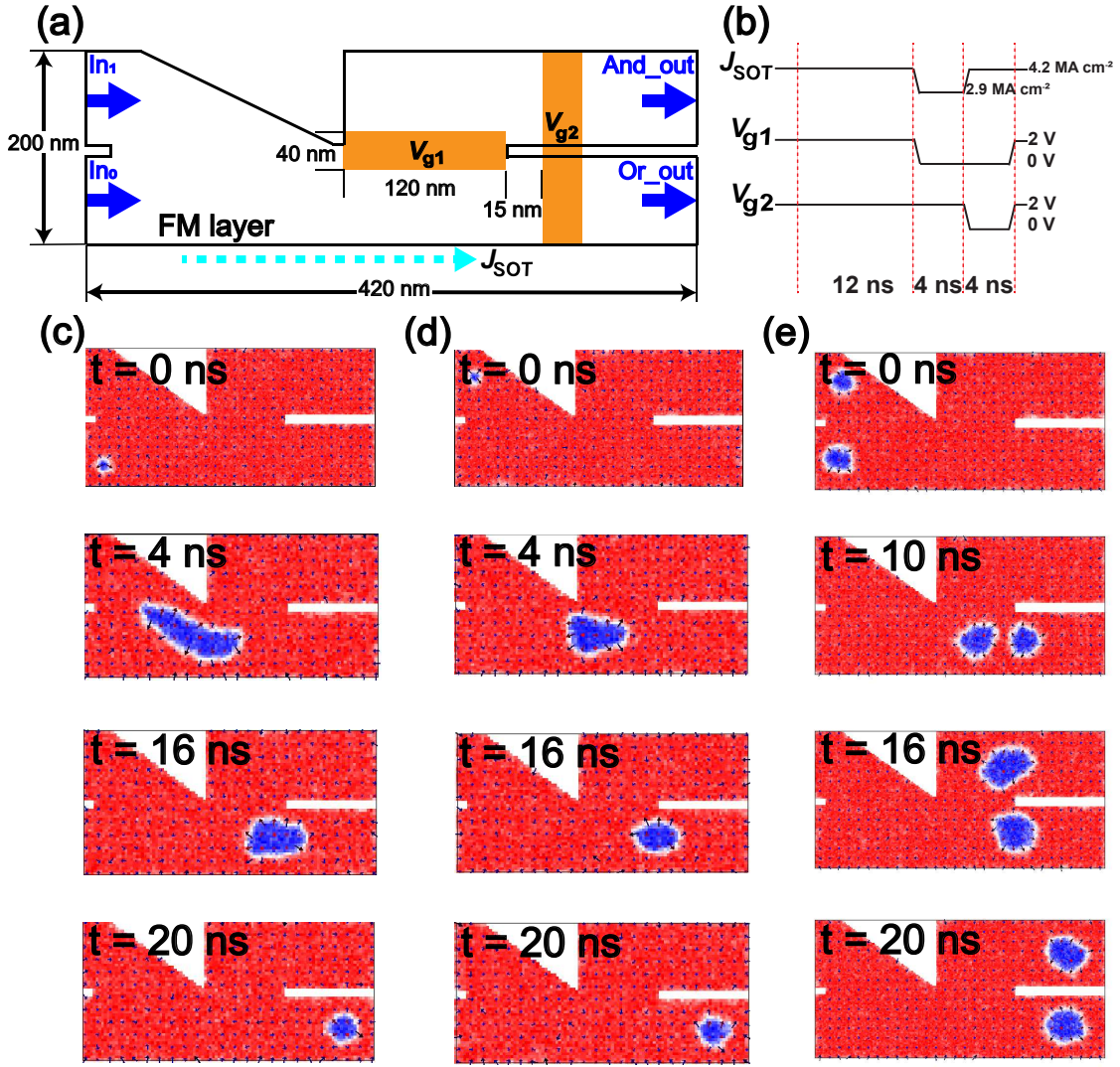

FIG. 6 Skyrmionic AND-OR logic device performed at $T$ = 250 K: (a) the device structure is larger than the device in Fig.1(a). (b) Profile of $J_{SOT}$ and electrode voltages. (c-e) Three cases of inputs and their corresponding simulation process and results.

## Appendix C: Simulation results of SOT driven skyrmion motion at the interface where DMI changes

We investigate SOT driven skyrmion motion at different interfaces where DMI changes. A constant SOT current $J_{SOT}$ = 4 MA cm$^{-2}$ is applied along +$x$ direction to drive skyrmion motion. As shown in Fig. 7(a) where the interface is exactly along $y$ direction, skyrmion deflection is along +$y$ direction when $D_1 \leqslant D_2$. However, skyrmions will stop at the interface when $D_1 > D_2$. In contrast, in Fig. 7(b), we find that an oblique interface can guide skyrmions along –$y$ direction, which can be utilized to perform the MUX function. We further evaluate the effect of different DMI on the potential energy. We initialize a skyrmion at the left side of the nanotrack shown in Fig. 7(c). Along this nanotrack, the DMI strength $D$ increases from 3.2 mJ m$^{-2}$ at the left side to 3.8 mJ m$^{-2}$ at the right side. Due to the DMI gradient, the skyrmion will move freely to the right side and the energy is measured at each position. We can see that the potential energy in the region with a lower DMI is higher, which can be used to guide the motion of skyrmions. Besides, the energy difference between two regions is nearly proportional to the DMI difference. By enlarging the DMI difference between two adjacent regions, we can regulate the skyrmion in a more stable manner, and a higher current can be used to drive the skyrmion.

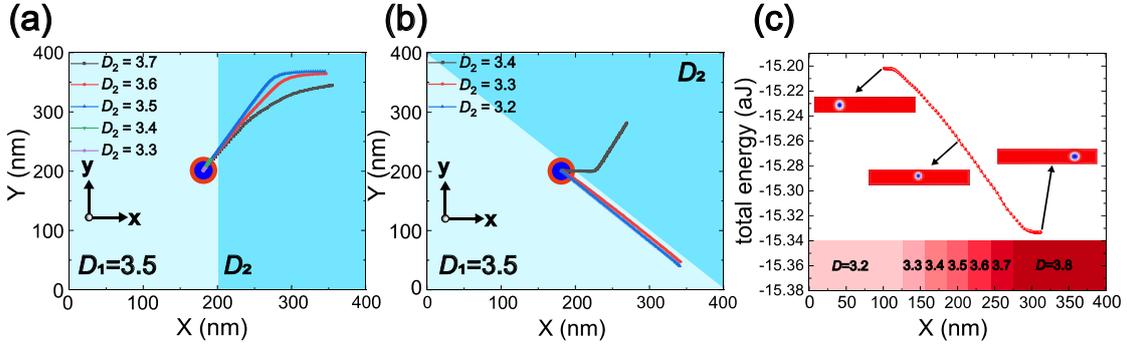

FIG. 7 (a-b) Trajectories of a SOT driven skyrmion at different $D_2$ when an $y$-direction or oblique interface is introduced. $D_1$ = 3.5 mJ m$^{-2}$ and $J_{SOT}$ = 4 MA cm$^{-2}$. (c) Effect of DMI on the energy of a nanotrack system. DMI strength increases from 3.2 mJ m$^{-2}$ at the left side to 3.8 mJ m$^{-2}$ at the right side, generating a force to drive the skyrmion along +$x$ direction. At each position, the energy of the system is recorded.

**Appendix D: Energy Consumption Evaluation**

Energy consumed in the AND-OR and MUX logic operation is calculated by:

$$W = J^2 \cdot S_{HM} \cdot \rho \cdot l \cdot T,$$

where $J$ is the SOT current density flowing in the HM layer, $S_{HM}$ is the $y$-$z$ cross area of HM layer, $\rho$ is the resistivity of the HM layer [59], $l$ is the length of the HM layer in $x$ direction and $T$ is the time duration. According to the time diagram shown in Fig. 1(b), different current density $J_i$ (i = 1, 2, 3) is applied for different time duration $T_i$. The corresponding parameters are listed in the TABLE I. Please note here that the capacitive charging/discharging energy of the VCMA gates is not included in the energy analysis, as this energy is generally negligible compared to the ohmic losses.

Therefore, the energy consumption for the AND-OR and MUX logic gates (Fig. 4(b)) are 7.19 fJ and 7.29 fJ respectively, while their operation speed is 8 ns and 15 ns, respectively. Following this method, the power consumption for the 3-bit stochastic multiplier and adder are evaluated as 1.264 pJ and 1.224 pJ.

**TABLE I. HM layer parameters for AND logic operation**

|  | $J_i$ (MA cm$^{-2}$) | $S_{HM}$ (nm$^2$) | $\rho$ ($\Omega \cdot$m) | $l$ (nm) | $T_i$ (ns) |
|---|---|---|---|---|---|
| i = 1 | 6 | 480 | $1.8 \times 10^{-6}$ | 2400 | 3 |
| i = 2 | 10 | 480 | $1.8 \times 10^{-6}$ | 2400 | 2 |
| i = 3 | 6 | 480 | $1.8 \times 10^{-6}$ | 2400 | 3 |

**Appendix E: Simulation of the skyrmionic logic devices in SAF**

We also adopt trilayer SAF systems to improve the performance of our devices [34]. The trilayer SAF system consists of two FM layers that are antiferromagnetically-exchange coupled through a non-ferromagnetic spacer. The initial magnetization of the top FM layer is along +z, while the lower FM layer is along -z. Because there is no skyrmion Hall Effect in such a system, we adjust the location of the electrodes based on our proposed MUX device as shown in Fig. 4, while the size remains unchanged. For the skyrmionic AND-OR logic device, we slightly extend the size along the *x* direction (200 nm to 240 nm) to get more space for skyrmion motion. The simulation results show that skyrmions in the SAF system can move about 2 times faster than the skyrmions in the FM system given the same energy consumption. Similarly, the performance of the MUX logic gate is optimized as well in the SAF system. Though the DMI strength used in our simulations, which is also adopted in Ref. [34] and Ref. [35], is higher than that obtained in experiments [60], it could be further optimized in the long term. Besides, by increasing the layers of a SAF system, the DMI required to stabilize skyrmions below 10 nm at room temperature can be lower than 1 mJ m$^{-2}$ [33], which further indicates the feasibility of our method. Therefore, our simulations in SAF show that our scheme corresponds to a practical material system, i.e. Co-based synthetic antiferromagnets. It will also be interesting to demonstrate our schemes in ferrimagnets, another promising material system, where F. Büttner and L. Caretta et al. have done much pioneer work [30,32].

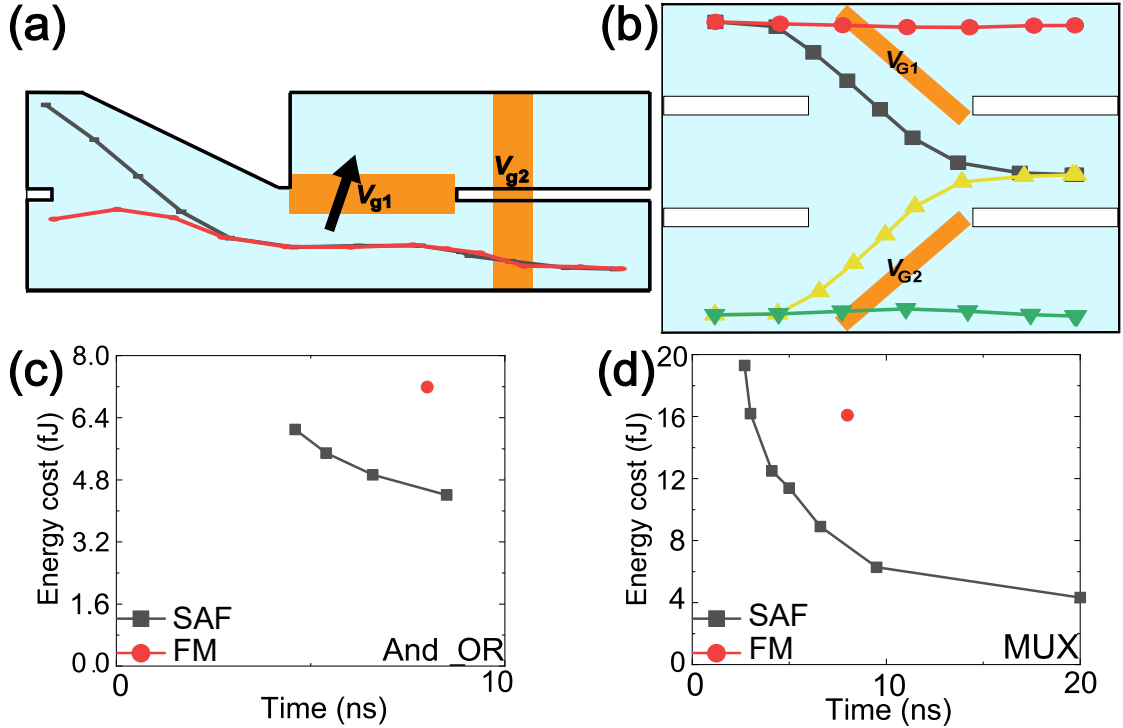

FIG. 8 (a-b) Trajectories of skyrmions in the skyrmionic logic device in SAF. (c-d) Comparison between the energy cost and time delay of our skyrmionic logic devices in SAF and in FM.